\documentstyle[aps]{revtex}
\begin{document}
\draft
\title{Formation of two-dimensional weak localization \\ in conducting Langmuir-Blodgett films}
\author{Yasuo Ishizaki$^\ast$, Mitsuru Izumi$^\ast$, Hitoshi Ohnuki$^\ast$, Krystyna Kalita-Lipinska$^\ast$, \\ 
Tatsuro Imakubo$^\dagger$, and Keiji Kobayashi$^\dagger$}
\address{$^\ast${\it Laboratory of Applied Physics, Tokyo University of Mercantile Marine, \\ 
2-1-6 Etchu-jima, Koto-ku, Tokyo 135-8533, Japan\/} \\ 
$^\dagger${\it Department of Chemistry, Graduate School of Arts and Sciences, The University of Tokyo, \\ 3-8-1 Komaba, Meguro-ku, Tokyo 153-8902, Japan\/}}
\date{\today}
\maketitle
\begin{abstract}
We report the magnetotransport properties up to 7 T in the organic highly conducting Langmuir-Blodgett(LB) films formed by a molecular association of the electroactive donor molecule bis(ethylendioxy)tetrathiafulvalene (BEDO-TTF) and stearic acid CH$_3$(CH$_2$)$_{16}$COOH.
We show the logarithmic decrease of dc conductivity and the negative transverse magnetoresistance at low temperature.
They are interpreted in the weak localization of two-dimensional (2D) electronic system based on the homogeneous conducting layer with the molecular size thickness of BEDO-TTF.
The electronic length with phase memory is given at the mesoscopic scale, which provides for the first time evidence of the 2D coherent charge transport in the conducting LB films.

\end{abstract}

The conducting $\pi$-charge-transfer salts of the electroactive molecule bis(ethylenedithio)tetrathiafulvalene (ET) have opened a frontier of material sciences like organic superconductivity~\cite{1,2,3,4}.
In the crystalline form, the ET-type donor molecules packed into layers separated by layers with a counter anion X$^-$ provide a two-dimensional (2D) electronic system~\cite{2,3,4} as the chemical composition (ET)$_2^+$X$^-$.
In this context, the conducting thin-films fabrication with ET-type donor molecular layers is crucial for potential applications in nano-scale molecular electronics~\cite{5,6}.
The Langmuir-Blodgett (LB) technique provides a layer by layer deposition according to the transfer of the molecular film compressed at the air-water interface onto solid substrate~\cite{5}.
Therefore, the LB technique is a preferable way to realize the conducting films with the ET-type donor molecule if the donor and acceptor molecules expanded on the water surface provide a stable conducting molecular film upon isothermal compression process.
Close intermolecular stack of the neighbors with partial charge transfer ultimately leads to a good metal film via $\pi$-molecular orbital overlap as in the crystalline form~\cite{6}.
Many works have been done to obtain a conducting LB film associated with the coherent carrier transport~\cite{7,8,9,10}. 
However, it was difficult to the LB film with high dc conductivity in a macroscopic scale.
Most essential obstacle is a nature of close packing of hydrophobic long alkyl chain~\cite{11}.
The long alkyl chain chemically incorporates with the ET-type donor molecule or counter anion, which increases the amphiphilic stability of the molecular film on the water surface.
It prevents the favorable close-packed organization for electroactive part bound to the long alkyl chain, resulting in the diffuse conduction like in semiconductor~\cite{11,12}.

In the present communication, we focus on the novel molecular association of an ET-type donor bis(ethylenedioxy)tetrathiafulvalene~\cite{9} (BEDO-TTF or BO, Fig. 1(a)) and stearic acid CH$_3$(CH$_2$)$_{16}$COOH (SA, Fig. 1(b)).
The BO and SA molecules form a highly conducting LB film.
This film structure shows that the above mentioned difficulty has been overcome.
The BO molecular layers alternate with the stearic acid layers in the Y-type LB film structure~\cite{5,9}.
The high dc conductivity up to 80 Scm$^{-1}$ at 300 K is originated from homogeneous conducting layers with the closed packed BO molecules. 
The existence of 2D weak carrier localization~\cite{13,14,15} is elucidated from the magnetotransport on a sheet of the Y-type LB film of BO and SA. 
A 2D weak localization itself is a mature physics and it verifies that the electronic system is in the regime of coherent carrier transport associated with the quantum interference effect.
The LB film of BO and SA is for the first time example of the conducting LB film that shows 2D coherent carrier transport at the mesoscopic scale.
The electronic length between dephasing collisions~\cite{13} grows up to 580 \AA \space at 1.7 K.
Thus, we report the quantum interference effect observed in the LB film, and thereby invite the study of conducting LB film to the mesoscopic physics~\cite{16}, as well as for potential application in quantum nanoelectronics~\cite{16,17}.

The LB film of BO and SA was obtained by so-called Y-type deposition following the similar preparation technique with the LB film of BO and behenic acid~\cite{9}.
Separate chloroform solutions of BO ($7.4\times10^{-4}$ mol/$\ell$) and SA ($8.22\times10^{-4}$ mol/$\ell$) molecules were prepared.
A mixed solution of BO and SA with molar concentration ratio 1:1 was spread on the pure water surface.
These molecules form an amphiphilic layer film BO-SA at the air-water interface upon isothermal compression process: the mono-molecular layer is made of SA, resulting in the lateral packing of long alkyl chain, and the BO molecules form a layer beneath the SA molecular layer~\cite{9,18}.
This is verified by a surface pressure/area isotherm that is similar to the isotherm of pure SA (Fig. 2 inset).
In the present layer on the water surface, partial charge transfer occurs between BO and SA to form a molecular association (BO)$_2^+$(RCOO$\cdots$HOOCR)$^-$, where R=CH$_3$(CH$_2$)$_{16}$~\cite{9,18,19}.
The counter anion is assigned as a COO$^-$ group of the SA~\cite{19,20}.
With keeping a surface pressure of 20 mN/m, the deposition starts on the first withdrawal from the subphase with the sapphire substrate whose hydrophilic surface is vertical with respect to the BO-SA layer film at the air-water interface.
The BO-SA layers are laid down each time the substrate moves across the boundary, which is a Y-type deposition~\cite{5,7,9}.
The number of layers is equal to the number of depositions.
Thus, we prepare the LB film with a stacking manner as BO-SA/SA-BO/BO-SA/SA-BO/... starting from the launch on the sapphire substrate. 

Figure 2 exhibits the lamellar X-ray diffraction profile from the 15-layer LB film of BO-SA at 300 K. 
Sharp diffraction peaks are observed.
The existence of a well-ordered stacking periodicity 58 \AA \space gives twice the BO-SA layer thickness according to the Y-type deposition~\cite{5,9}. 
Therefore, the 15-layer film thickness is 58 \AA/2$\times$15=435 \AA.
The film thickness was estimated to be from 435 \AA \space to 725 \AA \space for three samples with different number of depositions.
A model of the molecular organization is shown in the inset of Fig. 2, which was obtained from the stacking periodicity 58 \AA, the molecular orientation determined by electron-spin resonance study~\cite{21}, and the in-plane X-ray diffraction study.
The layer of (BO)$_2^+$, which is shown in Fig. 2, has a close-packed organization, which assures the optimal overlap of $\pi$-type HOMOs (highest occupied molecular orbital) of the neighbors in the film plane~\cite{3,6}. 
The obtained layered structure is a prerequisite of 2D electrical transport as shown in the single crystal (ET)$_2^+$X$^-$ ~\cite{2,3,4,11}.

The four-probe method was applied with 1.0 $\mu$A current flow to measure the dc electrical conductivity and magnetotransport.
Prior to the deposition of the LB film, four gold electrodes separated by 0.5 mm gaps were made by evaporation and chemical etching procedure~\cite{9} on the sapphire substrate.
The LB film was deposited on the present electrode-evaporated substrate.
Measurements of the transverse magnetoresistance up to 7 T were carried out with the superconducting magnet (Oxford MagLab-System2000).
The condition of the Ohmic contact was verified with the current-voltage characteristic down to 1.7 K.
The sample was cooled slowly (0.4 K/min.) to avoid irreversible jumps in conductivity.

Figure 3 shows the temperature variation of the film conductance {\it G\/} of the 21-layer LB film of BO-SA. This shows a substantial metallic character for {\it T\/}$>$120 K and {\it G\/} initiates to decrease logarithmically at least for {\it T\/}$<$40 K.
The conductance {\it G\/} is defined as ${\mit \sigma\/}{\mit d\/}$, where ${\mit \sigma\/}$ and ${\mit d\/}$ are the conductivity and the film thickness, e.g., ${\mit d\/}$ = 609 \AA \space for the 21-layer LB film, respectively.
In the present stage, we do not exclude the layer thickness of SA which is an insulator.
The inset of Fig. 3 shows the transverse magnetoresistance ${\mit \Delta\/}\rho/\rho_0=[\rho(B)-\rho(0)]/\rho(0)$ as a function of magnetic field applied at 1.8 K. 
The low-field magnetoresistance ${\mit \Delta\/}\rho/\rho_0$ is highly anisotropic. There is a negative amplitude when the magnetic field is applied normal to the film plane (${\mit \theta\/}=90^\circ$).
On the other hand, the magnetoresistance in ${\mit \theta\/}=0^\circ$ is positive monotonic with increasing the magnetic field.
The negative part of ${\mit \Delta\/}\rho/\rho_0$ disappears as the magnetic field vector approaches the film plane.
The similar low-field magnetoresistance anisotropy has been reported in the single crystal ET-type charge-transfer salt (DMtTSF)$_2$ClO$_4$ ~\cite{22}. (DMtTSF)$_2$ClO$_4$ is metallic and the results has been interpreted in terms of 2D weak localization in (DMtTSF)$_2$ conducting layer possibly due to the disorder in the anion layer~\cite{22}.

In Fig. 4, we show the transverse magnetoconductance ${\mit \Delta\/}${\it G\/} (${\mit \theta\/}=90^\circ$) using the definition ${\mit \Delta\/}G=G(B,T)-G(B=0,T)$ at several fixed temperatures.
The data was taken on the 25-layer LB film of BO-SA.
${\mit \Delta\/}${\it G\/} is enhanced as a positive ${\mit \Delta\/}${\it G\/} with decreasing temperature as shown in Fig. 4. 
The field variation of ${\mit \Delta\/}${\it G\/} at 1.7 K closely resembles the form of negative ${\mit \Delta\/}\rho/\rho_0$ taken at 1.8 K in Fig. 3.
The observed logarithmic decrease in zero-field {\it G\/} and positive ${\mit \Delta\/}${\it G\/} are common feature in the conducting LB films of BO-SA. This result indicates the formation of weakly localized 2D electronic system~\cite{13,14,15}.

In the coherent regime of conduction, the carrier mean free path ${\mit \lambda\/}$ is longer than its wavelength ${\mit \lambda\/}_F$.
Under the presence of the static random weak potential, if the phase coherence time for inelastic scattering ${\mit \tau\/}_{\it in\/}$ is larger than the elastic scattering time ${\mit \tau\/}$, the carrier diffuses at a distance called as Thouless length $L_{Th}=(D{\mit \tau\/}_{in})^{1/2}$ between dephasing inelastic scattering, where {\it $D=v_F^2{\mit \tau\/}/2$\/} and {\it $v_F$\/} are the diffusion constant and the Fermi velocity, respectively.
${\mit \tau_{in}\/}$ is proportional to {\mit $T^{-p}$\/} in which ${\mit p\/}$ is an index depending on scattering details~\cite{13,14,15}.
The weak localization results from the constructive quantum interference of partially backscattered electron waves travelling back along the time reversal paths, when the elastic scattering due to weak random potential occurs many times between acts of inelastic scattering at low temperature.

For 2D thin film, we have the sheet conductance {\it $G_{2D}(T)=G_0+(p/2)(e^2/\hbar\pi^2)\ln(T/T_0)$\/} from the scaling theory~\cite{13}.
For 40 K $<$ {\it T\/} $<$ {\it T$_0$\/} $=$ 120 K at which {\it ${\mit \tau\/}_{in}$\/} is close to ${\mit \tau\/}$, the conductance gradually deviates from the weak localization as {\it ${\mit \tau\/}_{in}$\/} approaches to ${\mit \tau\/}$.
Fitting {\it $G_{2D}(T)$\/} to the low-{\it T\/} data of {\it G\/} below 40 K in Fig. 3 reveals the conductance due to 2D weak localization to be temperature dependent: ${\it p\/}=1.4$ (${\it T_{\rm 0\/}\/}=120$ K). 
Here, in the fitting procedure on the sheet conductance {\it $G_{2D}(T)$\/}, we have excluded the thickness of the insulating SA layer (see Fig. 2 inset) from the estimation of the film thickness ${\mit d\/}$. 
The obtained value of ${\mit p\/}$ is not far from ${\mit p\/}=1$ observed in pregraphitic carbon fibers~\cite{23} and ${\mit p\/}=1.09$-1.29 for a metallic phase of Si/SiGe quantum wells~\cite{24}.
These values of ${\mit p\/}$ are close to the expected value ${\mit p\/}=1$ for dephasing by quasielastic electron-electron scattering~\cite{24,25}.
It is worth noting that the appearance of the weak localization at relative high temperatures has been reported in the graphite intercalation compounds~\cite{26} and some carbon nanotubes~\cite{27}. 

Below 40 K, as shown in Fig. 3, the carrier localization becomes dominant.
The elastic scattering loops leading to a weak localization presumably lie in 2D planes organized by an array of (BO)$_2^+$ molecules.
Externally applied magnetic field disrupts both the coherent back-scattering and the constructive quantum interference~\cite{13,15,28}.
So it suppresses the localization and results in positive ${\mit \Delta\/}${\it G\/} values, i.e., negative magnetoresistance.
As shown in both Figs. 3 and 4, negative magnetoresistance at low magnetic field and positive magnetoresistance at high field were both observed.
Therefore, the temperature dependence of ${\mit \Delta\/}${\it G\/} in Fig. 4 is interpreted by using the two-component model~\cite{22} which includes two independent terms in conductance, namely the classical contribution and the one terms associated with the 2D weak carrier localization in the intra-layer direction.
They show up in the magnetoconductance in the form~\cite{13,22,28}:
\begin{eqnarray}
{\mit \Delta\/}G=G(B,T)-G(B=0,T)=A_1\frac{e^2}{2\pi^2\hbar}\left[\ln\left(\frac{B}{B_i}\right)+\psi\left(\frac{1}{2}+B_i/B\right)\right]-A_2B^{\alpha},
\end{eqnarray}
where {\it $B_i=\hbar/4eL_{Th}^2$\/}, {\it $A_1$\/}, {\it $A_2$\/}, ${\mit \alpha\/}$ denote parameters used in the fitting of the magnetoconductance data, ${\mit \psi\/}$ the digamma function.
To fit the calculation according to the above formula to the experimental magnetoconductance, we determine four independent parameters {\it $A_1$\/}, {\it $B_i$\/}, {\it $A_2$\/} and ${\mit \alpha\/}$ at given temperatures.
The phenomenological prefactore {\it $A_1$\/} describes additional mechanism such as anisotropy and otherwise the value is 1~\cite{15,24}.
The factor {\it $A_2$\/} is primarily related to the effective magnetoresistance mobility in the classical magnetotransport~\cite{29}.
The exponent of the field in the classical part generally follows a ${\it B^\alpha\/}$ law and ${\mit \alpha\/}$ strongly depends on the electronic structure and disorder induced by anion species in the present molecular based conducting materials~\cite{22}.
In practice, we used the fixed value of ${\mit \alpha\/}=0.9$ for the classical positive part in the second term of ${\mit \Delta\/}${\it G\/}, then calculated the other parameters. The accuracy for the value {\mit $\alpha$\/} was $0.9\pm0.1$ in the final refinement.
The value ${\mit \alpha\/}=0.9(1)$ may indicate that the classical part obeys the linear magnetoresistance which is due to a possible inhomogeneous electron-density distribution in the present 2D electronic system~\cite{30}.
The definitive origin of this classical part and anisotropy will be clarified by the future high field magnetotransport with reducing the effect of weak localization.

Figure 4 gives the calculated solid curve fitted for experimental magnetoconductance ${\mit \Delta\/}${\it G\/} at 1.7 K, 3.5 K, 5.1 K, 8.3 K and 16.7 K.
All numerical values are summarized with the errors in Table I.
The relevant scale size {\it $L_{Th}$\/} over which quantum interference is onset increases up to 580 \AA, so that such weak localization is progressively evident at the mesoscopic scale.
We plot the obtained values of {\it $B_i$\/} as a function of temperature in Fig. 5.
The {\it $B_i$\/} which is described as {\it $\hbar/4eD{\mit \tau\/}_{in}$\/}, implies the temperature variation of the phase coherence time~\cite{31}.
Assuming that ${\mit \tau_{in}\/}={\it aT^{-p}\/}$, we obtain ${\mit a\/}=0.026$ with ${\mit p\/}=1\pm0.04$ (solid line in Fig. 5).
The value ${\mit p\/}$ is in not far from ${\mit p\/}=1.4\pm0.1$ obtained from the logarithmic temperature variation of the conductance ${\mit G\/}$ below 40 K for another sample in Fig. 3.
We think the difference is in sample to sample variation.
We have no knowledge for the value of ${\mit D\/}$ in the present conducting LB films.
For example, ${\mit D\/}$ assumes the value of $6.3\times10^{-3}$ m$^2$/s for thin filmed Cu with the thickness {\mit $d=90$ \AA~\cite{32}, we obtain  ${\mit \tau_{in}\/}=2.6\times10^{-13}$ s at 5 K where ${\mit B_i\/}=0.1$ T in Fig. 5.
For the Cu film the elastic scattering time is $6.5\times10^{-15}$ s, which implies that electrons can diffuse enough during a characteristic time and the obtained {\it $L_{Th}$\/} length plays the role of a system size for an electron in an eigenstate between inelastic scattering.
As stated by Senz {\it et al.\/}~\cite{24}, the value {\it $A_1$\/} is a phenomenological parameter that describes additional mechanisms such as anisotropic scattering~\cite{13}, or scattering by the Maki-Thompson mechanism~\cite{33}.
The factor {\it $A_1$\/} in Table I grows rapidly and becomes close to 1 with decreasing temperature.
According to the perturbative scaling theory for the 2D electronic system with in-plane anisotropy\cite{13}, the value {\it $A_1$\/} may become less than 1.
The obtained value $A_1=0.8$ at 1.7 K is close to unity and its temperature variation given in Table I suggests the formation of the weakly localized 2D metallic sheet with reducing the in-plane conductivity anisotropy in the LB film\cite{13}.
For the precise nature of the carrier inelastic scattering mechanism, in-plane anisotropy and its temperature dependence, more extended low-temperature investigation is necessary.
From the localization viewpoint, we are dealing with 2D system because
{\mit $L_{Th} \gg d$\/}, where ${\mit d\/}$ is an effective thickness of the film.
We found ${\mit d\/}=20$ \AA \space from the conducting layer thickness of BO in Fig. 2.
And the values {\it L$_{Th}$\/} in Table I fulfill the above condition.

In summary, the conducting LB film of BO-SA provides a well-defined layered stacking and exhibits high dc conductivity among the previously reported LB films. 
We have studied the film magnetoconductance {\it ${\mit \Delta\/}$G\/} in association with {\it $\ln(T)$\/} dependence of the zero-field film conductance {\it G\/} at low temperature.
These results were interpreted in terms of the weakly localized 2D electronic system formed in the conducting BO donor layers.
Temperature variation of the positive magnetoconductance indicates the growth of the relevant scale size {\it $L_{Th}$\/} over which quantum interference increases up to 580 \AA \space at 1.7 K, so that such weak localization is evident at the mesoscopic scale.
The phase coherence time reveals the temperature variation which resembles with the behavior of pregraphitic carbon fibers or the metallic phase in Si/SiGe quantum wells.
The amplitude of the positive magnetoconductance exhibits the formation of 2D weak localization with a small conductivity anisotropy in the film plane.
The trace of the in-plane anisotropy may be correlated to the inhomogeneity of the quasi-linear magnetoconductance in the analysis by using the two-component model.
The existence of the weakly localized 2D-electronic system indicates that the electronic transport is essentially in a coherent transport regime, influenced by the effect of weak random potential possibly produced by a small disorder inside the 2D molecular stacks of BO molecules.
These results show that the novel organic conducting LB film of BO-SA is an intriguing candidature for further investigation to pursue the dephasing and quantum interference transport properties coming from the reduced dimensional network of HOMO with BO donor molecules.

It is a pleasure to thank P. Delhaes, V. M. Yartsev and B. Desbat for helpful discussions.
This work was supported in part by the Research Fellowships of the Japan Society for the Promotion of Science for Young Scientists (Y.I.), a Grant-in-Aid for Scientific Research from the Ministry of Education, Science and Culture of Japan (M.I. and H.O.) and the Canon Foundation in Europe (M.I.).

*Author to whom correspondence should be addressed. E-mail address: ishizaki@zairyo.phys.tosho-u.ac.jp


%
%
\begin{table}
\label{parameters}
\caption{Parameters found in the fitting procedure of the magnetoconductance. Resulted parameters from the fitting procedure for the data in Fig. 4 with the calculation of the mangetoconductance ${\mit \Delta\/}${\it G\/} (equation (1)) based on a two-component model [13,22,28]. {\it $B_i = \hbar/4eL_{Th}^2$\/}. {\it $L_{Th}$\/} is a Thouless length. {\it $A_1$\/} and {\it $A_2$\/} are scaling parameters used in the fitting of the magnetoconductance data.}
\begin{tabular}{ccccc}\hline
{\it T\/} & {\it $A_1$\/} & {\it $B_i$\/} & {\it $A_2$\/} & {\it $L_{Th}$\/} \\
( K ) & $\space$ & ( T ) & $\space$ & ( \AA \space ) \\ \hline
1.7 & 0.82(1) & 0.049(1) & 5.51(3) $\times$ 10$^{-6}$ & 580(2) \\
3.5 & 0.86(1) & 0.087(1) & 4.76(4) $\times$ 10$^{-6}$ & 443(2) \\
5.1 & 0.70(1) & 0.115(1) & 1.67(5) $\times$ 10$^{-6}$ & 383(2) \\
8.3 & 0.63(2) & 0.205(2) & 0.93(7) $\times$ 10$^{-6}$ & 285(1) \\
16.7 & 0.33(2) & 0.475(3) & 0.40(8) $\times$ 10$^{-6}$ & 186(7) \\ \hline
\end{tabular}
\end{table}

%
%
\begin{figure}
\caption{Structures of BEDO-TTF (BO) and stearic acid (SA)}
\label{molecule}
\end{figure}

\begin{figure}
\caption{X-ray diffraction profile of the 15-layer LB film of 1:1 mixture of BO and SA. The stacking periodicity of the Y-type LB film is 58 \AA. Upper inset, surface pressure/area ($\pi$-A) isotherms of pure SA (open circles) and the 1:1 mixture of BO and SA molecules (closed circles) on a pure water subphase at 292 K. The horizontal axis shows the area per SA molecule. A close-packed model of the molecular organization of BO and SA molecules is shown in the lower inset. Thanks to the Y type of sequence [5], the LB film is organized by an array of the conducting (BO)$_2^+$ layers with a thickness of 20 \AA. Each conducting layer is sandwiched between insulating SA layers, leading to a Y-type layer unit (58 \AA \space in thickness), thereby the multilayered film is built up.}
\label{Xray}
\end{figure}

\begin{figure}
\caption{Temperature variation of the film conductance {\it G\/} for 21-layer LB film of 1:1 mixture of BO and SA on sapphire glass. Cooling rate is 0.4 K/min. Inset shows the transverse magnetoresistance {\it ${\mit \Delta\/}$$\rho/\rho$$_0$\/} for $\theta=0^\circ$ and $\theta=90^\circ$ at 1.8 K.}
\label{tempscan}
\end{figure}

\begin{figure}
\caption{Transverse magnetoconductance ${\mit \Delta\/}${\it $G=G(B, T)-G(B=0,T)$\/} for the 25-layer LB film of 1:1 mixture of BO and SA at 1.7, 3.5, 5.1, 8.3 and 16.7 K. Solid curves correspond to the calculated magnetoconductance based on a two-component model (equation (1)) including the contribution from the formation of the weakly localized 2D electronic system.}
\label{fieldscan}
\end{figure}

\begin{figure}
\caption{Temperature variation of {\it $B_{i}$\/} for the 25-layer LB film of 1:1 mixture of BO and SA.} 
\label{Thouless}
\end{figure}

\end{document}